# Quaternions in University-Level Physics Considering Special Relativity

Martin Erik Horn

University of Potsdam, Physics Education Research Group,
Am Neuen Palais 10, 14469 Potsdam, Germany
E-Mail: marhorn@rz.uni-potsdam.de

**Abstract**
As an expansion of complex numbers, the quaternions show close relations to numerous physically fundamental concepts (e.g. Pauli Matrices). In spite of that, the didactic potential provided by quaternion interrelationships in formulating physical laws are hardly regarded in the current physics curriculum. In particular, many approaches emerge that are useful in conveying the unity of seemingly distinct theories in a didactically convincing manner.
This will be exemplified with the help of special relativity. The diverse examples of spatial and space-time rotations are merged into a didactic whole by introducing quaternion transformations and comparing them to the representation using rotation matrices common in physics books.

**Contents**
1. Quaternions, a brief introduction
2. What are quaternions?
3. Didactic approaches to representing quaternions
4. Quaternions in university-level physics
5. Special relativity
6. Prospects
7. Bibliography
8. Appendix

## 1. Quaternions, a brief introduction

As an expansion of complex numbers quaternions present a foundation for a mathematically simple representation of rotations. Thus it is not surprising that especially in the field of software development the rotation of virtual structures in general is accomplished using the mathematics of quaternions rather than rotation matrices because of their formally easy implementation. This does not only save time in calculating, but is easily comprehensible and over all can be implemented without difficulty once the formalism is understood.

Thus it is worthwhile to further examine the didactic potential of mathematically representing physical processes using quaternions.

## 2. What are quaternions?

A quaternion q, technically speaking, is a four tuple with a one-dimensional, scalar part $q_o$ and a three-dimensional vector part $q_1 \cdot \mathbf{i} + q_2 \cdot \mathbf{j} + q_3 \cdot \mathbf{k}$ [2]. The basis vectors **i**, **j** and **k** are subject to the rules of multiplication

$$\mathbf{i}^2 = \mathbf{j}^2 = \mathbf{k}^2 = -1$$
$$\mathbf{i} \cdot \mathbf{j} = -\mathbf{j} \cdot \mathbf{i} = \mathbf{k}$$
$$\mathbf{j} \cdot \mathbf{k} = -\mathbf{k} \cdot \mathbf{j} = \mathbf{i}$$
$$\mathbf{k} \cdot \mathbf{i} = -\mathbf{i} \cdot \mathbf{k} = \mathbf{j}$$

established by William R. Hamilton on October 16, 1843 [1] and are responsible for significant properties of quaternions due to their anti-communicative structure. The representation of a rotation with the help of quaternions is generated as a result of the identification of the position vector x with the quaternion four tuple (0, x), whereas the rotation around the angle Θ is performed by a unity quaternion

$$q = (\cos \Theta/2 + \mathbf{u} \cdot \sin \Theta/2)$$

according to $(0, x') = q \cdot (0, x) \cdot q^*$.

The newly created position vector x' can be read easily from (0, x'), whereas the rotation axis corresponds to the unity quaternion (0, u) with $u = u_1 \cdot \mathbf{i} + u_2 \cdot \mathbf{j} + u_3 \cdot \mathbf{k}$ [2]. Specifically remarkable is the relationship between the rotation angle Θ and the quaternion angle Θ/2, which is also physically reflected in the spin with a rotational symmetry of 4π instead of the actually expected value of 2π. This astounding doubling of the angle can be attributed to the fact that two subsequent reflections at two planes, that are inclined toward each other at an angle of Θ/2, equal a rotation around the angle Θ.

## 3. Didactic approaches for representing quaternions

Numerous attempts exist in trying to visually represent the rotational symmetry of 720° or 4 π. Thus Richard Feynman at times successfully demonstrated in his lectures (see for example sequence of pictures in [4]) that the original state of a rotating marked cup is recovered only when rotated twice. If the object is rotated once at 360° the arm of the presenter remains in a topologically contorted state and to a certain extent is inextricable in this demonstrational experiment, which goes back to Dirac.

This inextricableness has to be considered explicitly for the representation of a quaternion multiplication and is the product of a changed algebraic sign of the quaternion state variable. Conway and Guy [3] realized this with their quaternion machine in that they employed an additional degree of freedom in the form of an indicator of an algebraic sign (twisted tie, see picture 1).

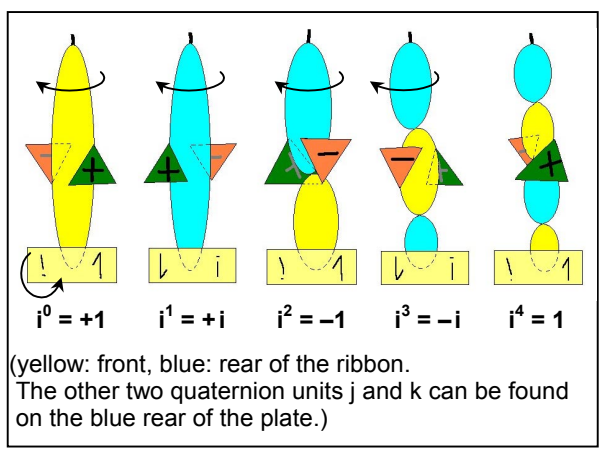

*Picture 1: Quaternion machine according to [3]. Shown is the fourfold multiplication by i.*

In accordance with an idea in [5] the difference between the original position of an object and the object rotated by 360° can be demonstrated by a simple construction. In picture 2 this was accomplished with the help of a square plate that was fastened to the corners of a box with rubber bands. If one turns the plate once, then the bands twist to the point where they are not able to be disentangled. Yet if the plate is turned twice and the rubber bands are consequently more twisted, they can however be disentangled with some patience.

## 4. Quaternions in university-level physics

If students are confronted with unfamiliar quantum mechanical phenomena such as the spin for the first time in their studies, this confrontation is usually accompanied by a failure of mathematical methods that had been effectively applied previously. They are taught that new physics require new mathematical instruments such as for example the Pauli matrices

$$\sigma_x^2 = \sigma_y^2 = \sigma_z^2 = 1$$
$$\sigma_x \sigma_y = -\sigma_y \sigma_x = i\sigma_z$$
$$\sigma_y \sigma_z = -\sigma_z \sigma_y = i\sigma_x$$
$$\sigma_z \sigma_x = -\sigma_x \sigma_z = i\sigma_y$$

This didactic approach to conveying new physical concepts in a practically revolutionary manner and by way of simultaneous concept changes in multiple fields, certainly corresponds with the significance that is ascribed to the modern quantum mechanics in the history of the development of modern physics.

This paper however will focus on the question of whether it would me more beneficial for an effective learning process to defuse this change of concepts represented simultaneously in a physical and mathematical manner by providing the students with new mathematical resources in advance. In the physical sense the concept of the quaternion (thus also the Pauli matrices) is actually nothing new and its main purpose allegedly serves to provide a more convenient description of an already known physical process, i.e. rotation.

It is therefore useful to discuss the mathematics of quaternions in connection with special relativity. The benefits are numerous:

- The mathematical formalism is more readily understood if it is linked to known physical phenomenon.
- The students are able to take a meta-conceptual standpoint when discussing the theory of relativity since they are familiar with a variety of mathematical approaches.
- A clear demonstration of special relativity is possible only by way of rotation matrices under consideration of one spatial coordinate, as this is the case in nearly all introductory physics schoolbooks. It is easier to formulate the Lorentz transformations without these restrictions, by using quaternions.
- Numerous other problems can be elegantly explained with the help of quaternions.

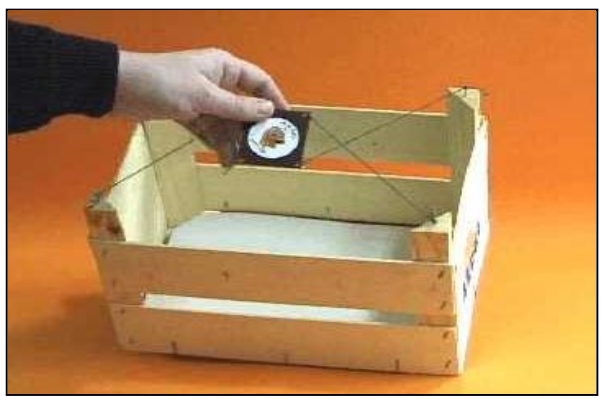

*Picture 2: Simple experiment to demonstrate the difference between rotations of 2π and 4π.*

## 5. Special relativity

In special relativity spacetime consists of three space dimensions x, y, and z as well as one of a time dimension ct. As the speed of light is constant, the four-dimensional distance between two events is equal in all inertial systems. A transformation between two inertial systems is thus considered an active rotation of a four vector

$$\vec{r} = \begin{pmatrix} ct \\ i \cdot x \\ i \cdot y \\ i \cdot z \end{pmatrix}$$

around the origin of the coordinate system in so far as translations are disregarded as physically uninteresting parallel shifts [6].

### - Representation using rotation matrices

A two-dimensional spatial rotation around the x-axis maps the coordinates onto each other according to

$$\vec{r}' = \begin{pmatrix} 1 & 0 & 0 & 0 \\ 0 & 1 & 0 & 0 \\ 0 & 0 & \cos\alpha & \sin\alpha \\ 0 & 0 & -\sin\alpha & \cos\alpha \end{pmatrix} \cdot \vec{r}$$

Naturally a real rotation angle $\alpha$ needs to be selected so that the dimension qualities do not change.

If one then shifts to a two-dimensional space-time rotation, the following rotation matrix is useful in its representation:

$$\vec{r}' = \begin{pmatrix} \cos\varphi & \sin\varphi & 0 & 0 \\ -\sin\varphi & \cos\varphi & 0 & 0 \\ 0 & 0 & 1 & 0 \\ 0 & 0 & 0 & 1 \end{pmatrix} \cdot \vec{r}$$

This is a Lorentz transformation as long as the angle of rotation is imaginary and is identified with the speed of transformation

$$v = i \cdot c \cdot \tan\varphi$$

according to [7].

Two aspects become apparent:

1) The structure of the matrix has changed. Solely space and space-time rotations are effectively differentiated based on the matrix positions that they take on.
2) The domains of definition of the angles need to be selected accordingly. A complex or a purely imaginary angle $\alpha$ leads to an unphysical situation as would a mixed-complex or purely real $\varphi$.

On the one hand, the illustration with the help of matrices has didactic benefits:

→ In order to correctly interpret the theoretical representation, students have to use physical arguments.

On the other hand, this can present didactic as well as aesthetic problems:

→ A theory that allows non-physical solutions can hardly be considered as complete.

From the viewpoint of special relativity it thus makes didactic sense to provide the students with theoretical resources with which the interpretation problems described above may be considered from another point of view. This approach will be discussed in the following with the representation of special relativistic transformations using quaternions.

### - Representation using quaternions

The Lorentz transformation was already formulated by both Einstein [8] and Dirac [9] using quaternions. However, their general representation is so abstract that an introduction at this level seems hopeless. The fundamental associations however can be reduced to a point didactically so that they can be instituted at the introductory level.

The four vector is from now on defined with the help of the quaternion units $\vec{i}, \vec{j}$ und $\vec{k}$:

$$R = ct + \vec{i} \cdot ix + \vec{j} \cdot iy + \vec{k} \cdot iz$$

The space rotation about the x-axis with angle $\alpha$ described on the left is produced by the quaternion

$$Q = \cos\frac{\alpha}{2} - \vec{i} \cdot \sin\frac{\alpha}{2} \qquad \alpha \in \mathbf{R}$$

As is typically the case for a quaternion illustration of a rotation, the transformation equation is as the following:

$$\begin{aligned} R' &= Q\, R\, Q^* \\ &= ct + \vec{i} \cdot ix \\ &\quad + \vec{j} \cdot (iy\cos\alpha + iz\sin\alpha) \\ &\quad + \vec{k} \cdot (-iy\sin\alpha + iz\cos\alpha) \end{aligned}$$

The new coordinates can be discerned from the brackets without difficulty and as expected equal:

$$\begin{aligned} y' &= y\cos\alpha + z\sin\alpha \\ z' &= -y\sin\alpha + z\cos\alpha \end{aligned}$$

whereas the angle $\alpha$ is real. As a result, complex conjugate quaternions describe an inverse rotation:

$$\begin{aligned} Q^* &= \cos\frac{\alpha}{2} + \vec{i} \cdot \sin\frac{\alpha}{2} \\ &= \cos\frac{-\alpha}{2} - \vec{i} \cdot \sin\frac{-\alpha}{2} = Q^{-1} \end{aligned}$$

A space-time rotation is obtained in the case of this quaternion approach by

$$Q = \cos\frac{\varphi}{2} - \vec{i}\cdot\sin\frac{\varphi}{2} \quad (\varphi \text{ purely imaginary})$$

Since with this choice the complex conjugate quaternion

$$Q^* = \cos\frac{-\varphi}{2} + \vec{i}\cdot\sin\frac{-\varphi}{2}$$
$$= \cos\frac{\varphi}{2} - \vec{i}\cdot\sin\frac{\varphi}{2} = Q$$

equals the original quaternion, a completely different transformational behavior develops:

$$R' = Q R Q^*$$
$$= (ct\cos\varphi + ix\sin\varphi)$$
$$+ \vec{i}\cdot(-ct\sin\varphi + ix\cos\varphi)$$
$$+ \vec{j}\cdot iy$$
$$+ \vec{k}\cdot iz$$

As a result this choice of the angle corresponds to a Lorentz transformation

$$ct' = ct\cos\varphi + ix\sin\varphi$$
$$x' = i\,ct\sin\varphi + x\cos\varphi$$
$$y' = y$$
$$z' = z$$

as far as the speed of transformation is again identified with

$$v = i\,c\,\tan\varphi \quad .$$

Consequently, physical-structural as well as didactic benefits develop:

1) The structure of the mathematical object Q remains unchanged.
2) The extension of the domain of definition of the rotation angle no longer leads to unphysical situations.

This leads to the fact that a discussion concerning the theory behind the theories and thus to a conceptionally all-encompassing analysis of special relativity by the students becomes possible.

**- Continued unification**

The next question posed is what happens when a mixed-complex rotation angle is selected. The quaternion approach can induce an encompassing understanding of physics and build a bridge back to the representation of matrices by posing this important question.

Thus

$$Q = \cos\frac{\alpha+\varphi}{2} - \vec{i}\cdot\sin\frac{\alpha+\varphi}{2}$$

The quaternion four vector transforms itself with this (see appendix) into

$$R' = Q R Q^*$$
$$= (ct\cos\varphi + ix\sin\varphi)$$
$$+ \vec{i}\cdot(-ct\sin\varphi + ix\cos\varphi)$$
$$+ \vec{j}\cdot(iy\cos\alpha + iz\sin\alpha)$$
$$+ \vec{k}\cdot(-iy\sin\alpha + iz\cos\alpha)$$

Moreover, the new coordinates can be discerned from the brackets:

$$ct' = ct\cos\varphi + ix\sin\varphi$$
$$x' = i\,ct\sin\varphi + x\cos\varphi$$
$$y' = y\cos\alpha + z\sin\alpha$$
$$z' = -y\sin\alpha + z\cos\alpha$$

First of all it is shown that the formalism introduced here classifies the real part of the angle automatically as a space rotation, whereas the imaginary part causes a space-time rotation. The necessary additional conditions for the matrix representation are not needed.

Secondly, the relevant matrix representation

$$\vec{r}' = \begin{pmatrix} \cos\varphi & \sin\varphi & 0 & 0 \\ -\sin\varphi & \cos\varphi & 0 & 0 \\ 0 & 0 & \cos\alpha & \sin\alpha \\ 0 & 0 & -\sin\alpha & \cos\alpha \end{pmatrix}\cdot\vec{r}$$

can be easily deduced from these equations without problems. It becomes immediately obvious that this rotation matrix is the product of both of the original matrices as expected.

**6. Prospects**

The approach presented here provides numerous possibilities for development. Thus one of the most impressive properties of the relativity theory is not only that Maxwell's equations are contained within it in their entirety, but also that the formulation of these equations can be presented more clearly. This is due to the fact that the equations possess a higher symmetry within the framework of special relativity.

Having important consequences for the learning process, the analysis of quaternion representations of other relativistic relationships should be a further theme of physics education research.

If one would like to moreover work through the general relativity theory didactically, then it is hardly possible to avoid that the interaction between group theoretic aspects, the Pauli matrices and the theory of relativistic phenomenon is researched in more detail.

A further objective emerges in the conceptual differentiation presented here in the text between the quaternion unit $\vec{i}$ and the imaginary unit i. Even though the Lorentz transformation was presented within the context of a quaternion representation, it is however not clear what didactic consequences a strictly octonion formulation can provide.

However, all of these possible dilations remain inadequate didactically if epistemological conditions, philosophical views, or, as is illustrated in [10], the current *Zeitgeist* of the surroundings within which the students and teachers are acting are not simultaneously considered. This will be briefly illustrated by three of the four discussed meta-principles of chapter 13 [10] (*Philosophy in Physics*) that have an influence on physical thinking, research and study.

### - The unity of nature

The meta-principle of the unity of the natural laws postulates, „*that all nature is amenable to the same kind of theoretical treatment*" [10]. Even though the description of special relativistic effects by means of rotation matrices as well as the description using quaternions both adhere to this meta-principle, it is possible to work through the differing stages of implementation of the meta-principle with the students, while discussing both approaches. Due to its structural density, the quaternion representation is without a doubt a more unified theory in comparison to the matrix representation.

### - The principle of plenitude

This principle, which is intensely discussed among scientists, states that „*anything which is not prohibited is compulsory*" [10]. The task of physics education research should be to enable the students not only to receptively follow the discussion, but also to actively find a convincing standpoint. The comparison of rotation matrices and quaternions allows didactic access to such a cognitive process by analyzing the properties of angles. Specifically the question of whether rotation matrices with complex angles are feasible potentially introduces some difficulty in following this principle.

The outcome of the discussion may be open and may lead to an inverse meta-principle as was indirectly formulated in section 5: If the potential outcomes do not appear in nature as they were described theoretically, then the theoretical description might have gaps or is at least inappropriate. Regardless of whether different physicists or students of physics will invariably have different opinions concerning this question, the impact this meta-principle has had on the development of physics should be pointed out frequently. Dirac postulated the existence of positrons and magnetic monopoles based solely on this principle.

### - The principle of mathematical beauty

The (mathematical) beauty of a physical theory is actually not a comprehensible property, however it repeatedly makes its appearance in fundamental discussions comparing different theories. Bearing special relativity in mind, arguments for and against the beauty of the approaches introduced here will resurface. Such appraisals of the theoretic-structural beauty of a physical-mathematical representation orient themselves less on abstract concepts and rather on practical categories like convenience in application, i.e. *convenient theories* or complexity of structure, i.e. *complicated theories* [10]. Ultimately, the intention behind the request for (mathematical) beauty is to work out a possibly transparent and concise theoretical configuration of a physical thought construct, i.e. a very didactical intention.

Similarly the other meta-principles can be understood as didactic central themes that are able to be implemented with the support of special relativity.

### - Special thanks

I would like to thank Ralf Bürger (http://www.uni-potsdam.de/u/physik/didaktik/homepage/bue/rbuerger1.htm ) and all other colleagues of the physics education research group at the University of Potsdam for the help in developing the materials for the experiments.

## 8. Appendix

For any angle $\psi = \alpha + \varphi$ ($\alpha$ is real, $\varphi$ is imaginary and thus $\psi \in \mathbf{C}$) with the help of

$$Q = \cos\frac{\psi}{2} - \vec{i} \cdot \sin\frac{\psi}{2}$$

the following transformation is developed:

$$\begin{aligned}
R' &= Q R Q^* \\
&= \left(\cos\frac{\psi}{2} - \vec{i}\cdot\sin\frac{\psi}{2}\right)\cdot\left(ct + \vec{i}\cdot ix + \vec{j}\cdot iy + \vec{k}\cdot iz\right)\cdot\left(\cos\frac{\psi^*}{2} + \vec{i}\cdot\sin\frac{\psi^*}{2}\right) \\
&= \left[\left(ct\cos\frac{\psi}{2} + ix\sin\frac{\psi}{2}\right) + \vec{i}\cdot\left(ix\cos\frac{\psi}{2} - ct\sin\frac{\psi}{2}\right) + \vec{j}\cdot\left(iy\cos\frac{\psi}{2} + iz\sin\frac{\psi}{2}\right)\right. \\
&\qquad \left. + \vec{k}\cdot\left(iz\cos\frac{\psi}{2} - iy\sin\frac{\psi}{2}\right)\right]\cdot\left(\cos\frac{\psi^*}{2} + \vec{i}\cdot\sin\frac{\psi^*}{2}\right) \\
&= \left(ct\cdot\left(\cos\frac{\psi}{2}\cos\frac{\psi^*}{2} + \sin\frac{\psi}{2}\sin\frac{\psi^*}{2}\right) + ix\cdot\left(\sin\frac{\psi}{2}\cos\frac{\psi^*}{2} - \cos\frac{\psi}{2}\sin\frac{\psi^*}{2}\right)\right) \\
&\quad + \vec{i}\cdot\left(ix\cdot\left(\cos\frac{\psi}{2}\cos\frac{\psi^*}{2} + \sin\frac{\psi}{2}\sin\frac{\psi^*}{2}\right) - ct\cdot\left(\sin\frac{\psi}{2}\cos\frac{\psi^*}{2} - \cos\frac{\psi}{2}\sin\frac{\psi^*}{2}\right)\right) \\
&\quad + \vec{j}\cdot\left(iy\cdot\left(\cos\frac{\psi}{2}\cos\frac{\psi^*}{2} - \sin\frac{\psi}{2}\sin\frac{\psi^*}{2}\right) + iz\cdot\left(\sin\frac{\psi}{2}\cos\frac{\psi^*}{2} + \cos\frac{\psi}{2}\sin\frac{\psi^*}{2}\right)\right) \\
&\quad + \vec{k}\cdot\left(iz\cdot\left(\cos\frac{\psi}{2}\cos\frac{\psi^*}{2} - \sin\frac{\psi}{2}\sin\frac{\psi^*}{2}\right) - iy\cdot\left(\sin\frac{\psi}{2}\cos\frac{\psi^*}{2} + \cos\frac{\psi}{2}\sin\frac{\psi^*}{2}\right)\right) \\
&= \left(ct\cdot\left(\cos^2\frac{\varphi}{2} - \sin^2\frac{\varphi}{2}\right) + 2\cdot ix\cdot\sin\frac{\varphi}{2}\cos\frac{\varphi}{2}\right) \\
&\quad + \vec{i}\cdot\left(ix\cdot\left(\cos^2\frac{\varphi}{2} - \sin^2\frac{\varphi}{2}\right) - 2\cdot ct\cdot\sin\frac{\varphi}{2}\cos\frac{\varphi}{2}\right) \\
&\quad + \vec{j}\cdot\left(iy\cdot\left(\cos^2\frac{\alpha}{2} - \sin^2\frac{\alpha}{2}\right) + 2\cdot iz\cdot\sin\frac{\alpha}{2}\cos\frac{\alpha}{2}\right) \\
&\quad + \vec{k}\cdot\left(iz\cdot\left(\cos^2\frac{\alpha}{2} - \sin^2\frac{\alpha}{2}\right) - 2\cdot iy\cdot\sin\frac{\alpha}{2}\cos\frac{\alpha}{2}\right) \\
&= (ct\cdot\cos\varphi + ix\cdot\sin\varphi) + \vec{i}\cdot(ix\cdot\cos\varphi - ct\cdot\sin\varphi) \\
&\quad + \vec{j}\cdot(iy\cdot\cos\alpha + iz\cdot\sin\alpha) + \vec{k}\cdot(iz\cdot\cos\alpha - iy\cdot\sin\alpha)
\end{aligned}$$

This result gives the following transformation equations:

$$ct' = ct\cdot\cos\varphi + ix\cdot\sin\varphi \qquad ix' = ix\cdot\cos\varphi - ct\cdot\sin\varphi$$
$$iy' = iy\cdot\cos\alpha + iz\cdot\sin\alpha \qquad iz' = iz\cdot\cos\alpha - iy\cdot\sin\alpha$$

---